\documentclass[doublecol]{epl2} 
\usepackage{graphicx}
\usepackage{amssymb,amsmath}
\usepackage{array}
\usepackage{dcolumn}

\title{Real-space fluctuations of effective exchange integrals in high-$\bf T_c$
  cuprates} 
 
\author{S\'ebastien Petit  \and Marie-Bernadette Lepetit}
 
\institute{CRISMAT, ENSICAEN-CNRS UMR6508, 6~bd. Mar\'echal Juin, 14050 Caen, 
FRANCE}

\date{\today}

\abstract{
  We present ab-initio calculations of effective magnetic exchange, $J$, as
  well as Hubbard parameters ($t$, $U$ and $\delta \varepsilon$) as a function
  of the local distribution of doping atoms for the high-$T_c$ superconducting
  $\rm (Ca_xLa_{1-x})(Ba_{1.75-x}La_{0.25+x})Cu_3O_y$ family.  We found that
  both the exchange and the energies of the magnetic orbitals are strongly
  dependant on the local dopant distribution, both through the induced
  modification of the apical oxygen location and of the induced local
  electrostatic potential.  The $J$ real-space map, for a random distribution
  of dopants, positively compares with observed STS gap inhomogeneity
  maps. Similarly, the orbital energy fluctuations induce weak charge
  inhomogeneities on copper sites, that can be positively compared with the
  observed LDOS inhomogeneities. These results clearly support an extrinsic
  origin of both the gap inhomogeneities and LDOS.}

\pacs{74.62.Dh}{Effects of crystal defects, doping and substitution}
\pacs{75.30.Et}{ Exchange and superexchange interactions}
\pacs{71.10.Fd}{Lattice fermion models (Hubbard model, etc.)}
\begin{document}
\maketitle


Since their discovery by M\"uller in 1986~\cite{Muller86}, high-$T_c$
superconductors have been the subject of an extensive research activity, both
experimentally and theoretically. Recently, the question of the spatial
(in)homogeneities of the superconducting state has attracted a lot of
attention. Indeed, in the recent years, the improvement and development of the
tunneling microscopy techniques such as Scanning Tunneling Spectroscopy (STS)
have allowed the access to spatial imaging of quantities such as the density
of states and the superconducting gap. Using these techniques, nanoscale
spatial inhomogeneities of the gap amplitude were
observed~\cite{Cren00}. Strong fluctuations of the superconducting gap or even
suppression of the superconductivity was found within islands of typical size
of one nanometer in the $\rm Bi_2Sr_2CaCu_2O_{8+\delta}$ (Bi-2212) compound at
4.2K, that is far below the critical temperature~\cite{Pan01}.  The origin of
these gap inhomogeneities is quite controversial, however crucial since some
models link the high value of $T_c$ to electronic phases separation or
disorder.  The controversy opposes authors defending the idea that this is an
intrinsic feature of the superconducting state~\cite{Pan01,Pan02,Intrinseque},
to authors defending the idea of an extrinsic origin~\cite{Extrinseque},
linked to the chemical inhomogeneities. The first issue is directly related to
the models originating the high-$T_c$ superconductivity in electronic phases
separation~\cite{Phillips03}. Indeed, in such a case, the gap inhomogeneities
would be a pure electronic effect in an otherwise (chemically and
structurally) homogeneous system. The second issue emphasizes the role of the
always present chemical disorder resulting from the doping process out of the
Mott-Hubbard insulating parent system.

Despite the unknown origin of high-$T_c$ superconductivity it is widely
accepted that the antiferromagnetic correlations play an important role in its
onset, either as a competing phase~\cite{Tremblay08} or as a pairing
mechanism~\cite{Anderson04}. Some authors even claim a direct scaling of the
critical temperature with the effective magnetic exchange, $J$, between
adjacent copper atoms~\cite{Xino,Keren06}.  In any case $J$ plays a central
role in the superconductivity onset and thus its fluctuations as a function of
the chemical and structural inhomogeneities is expected to yield valuable new
insight on the origin (intrinsic or extrinsic) of the observed gap and $T_c$
spatial fluctuations.

The present work proposes to study the local fluctuations of the effective
exchange integral, $J$, as a function of the chemical and structural disorder
in the $\rm (Ca_xLa_{1-x})(Ba_{1.75-x}La_{0.25+x})Cu_3O_y$ (CLBLCO) copper
oxides family. Indeed, this family presents two determinant characteristics,
(i) the effect of hole doping and of chemical disorder is
decoupled~\footnote{A simple formal charges analysis shows that the doping in
  the CLBLCO family is independent of the amount of calcium, $x$, introduced
  in the compound~; the hole doping only depends on the oxygen concentration
  $y$. Two different copper oxide planes are crystallographically present in
  CLBLCO family~: the superconducting ones, associated with crystallographic
  Cu II sites, that do not see much fluctuations of their oxygen composition
  and the non superconducting ones, associated with Cu I sites, with a $y-6$
  oxygen composition.} (ii) structural informations on the local distortions
as a function of the chemical disorder are available~\cite{Struc}.


For this purpose we  used the CAS~\cite{CAS}+DDCI~\cite{DDCI} (Complete Active Space +
Difference Dedicated Configurations Interaction) quantum chemical spectroscopy
method on embedded fragments. Indeed, this method is presently the most
reliable and accurate one for the determination of effective magnetic
integrals. It  proved its efficiency on copper oxides
superconductors and predicted the $J$ values within experimental accuracy for
a large number of such compounds~\cite{Xino}.

The CAS+DDCI method is an exact diagonalization method within a selected set
of electronic configurations, specifically chosen so that to properly treat
(i) the strong electronic correlation within the set of magnetic orbitals and
(ii) the screening effects on the all magnetic configurations. The orbital
space is thus divided into three subsets~: the {\bf occupied orbitals} set
spanning orbitals always doubly-occupied in the reference configurations (the
CAS), the {\bf active orbitals} set for which all possible orbital occupations
and spins are found in the CAS configurations and the {\bf virtual orbitals}
set spanning orbitals always remaining empty in the CAS configurations. If the
active orbitals are defined as the magnetic orbitals of the copper atoms, the
CAS definition will thus insure the proper treatment (by exact
diagonalization) of the correlation effects within them. The screening effects
are then treated by adding to the diagonalization space all single and double
excitations, on all CAS configurations, that participate to the excitations
energies at the second order of the quasi-degenerate perturbation theory.
%

The embedded fragments are divided into a quantum part and a bath.  The
quantum part includes the magnetic atoms (copper atoms), the ligands mediating
the interaction as well as the first coordination shell of magnetic and
ligands fragments. The bath is built in order to reproduce the main effects of
the rest of the crystal on the quantum part, that is the Madelung potential
and the exclusion effects. The latter corresponds to the orthogonalisation of
the quantum fragment orbitals to the orbitals of the rest of the crystal. This
effect is usually taken into account through total ions pseudo
potentials~\cite{TIPS} (TIPs) located at the crystallographic positions of the
first coordination shell of the quantum fragment. The former is reproduced
using a set of renormalized charges.


The $\rm (Ca_xLa_{1-x})(Ba_{1.75-x}La_{0.25+x})Cu_3O_y$ compounds with $0.1
\le x \le 0.4$ and $6.8\le y \le 7.25$  are high temperature
double-layers superconductors belonging to the YBCO group. 
As in the layered YBCO structure, the CLBLCO compounds are built from two
different types of copper planes~: (i) two $\rm CuO_2$ superconducting layers
corresponding to the $\rm Cu\,II$ crystallographic sites (see
ref.~\cite{Struc}) separated by randomly distributed calcium and lanthanum
ions (named $\rm La_Y$), and (ii) basal $\rm Cu O_{y-6}$ layers, corresponding
to the crystallographic $\rm Cu\,I$ coppers. In the latter planes, the oxygen
atoms (named $\rm O_\alpha$) are randomly distributed along the Cu--Cu
bonds. Let us note that on the contrary to the YBCO compounds, no extended
$\rm CuO$ chains are observed in these $\rm Cu\,I$ layers.  The YBCO Ba sites
are randomly occupied with barium and lanthanum ions (named $\rm
La_{Ba}$). Associated with the apical oxygen atoms ($\rm O_c$), they form
layers in between the $\rm Cu\,II$ and $\rm Cu\,I$ planes.

From the above crystallographic data~\cite{Struc} one can spot four origins
for the chemical and structural disorder. (i) The random distribution of the
$\rm Ba$ and $\rm La_{Ba}$ ions in the YBCO Ba site. (ii) The in plane
displacements of the $\rm O_c$ (apical) and $\rm O_\alpha$ ($\rm Cu\,I$
layers) oxygen atoms as a function of the doping ion on the YBCO Ba
site. Indeed, the $\rm Ba^{2+}$ ions are $20\%$ larger than the $\rm La^{3+}$
ones. (iii) The random distribution of $\rm Ca$ and $\rm La_Y$ ions in the YBCO
Y site. (iv) The random distribution of the $\rm O_\alpha$ ions.  

It is usually
accepted that the exchange couplings are mainly sensitive to the geometry of
the coppers first coordination shells.  In the CLBLCO family this geometry is
controlled by the YBCO Ba sites occupation~\cite{Struc}.  In the present
work we thus focused on the effect of this degree of freedom, and averaged the
electrostatic effects induced by the $\rm O_\alpha$ positional disorder as
well as those induced by the $\rm Ca\,$/$\,\rm La_{Y}$ chemical disorder.

\begin{figure}[h] \centerline{
      \resizebox{8cm}{!}{\includegraphics{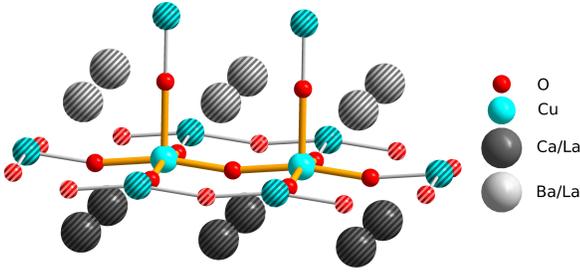}}}
\caption{(Color online) Schematic representation of the quantum part and TIPs
  (dashed) of the embedded fragments. The difference between the quantum part
 and the TIPs is emphasized by used of undashed/dashed atoms and large/small
 bond lines. }
\label{fig:frag}
\vspace*{-1eM}
\end{figure}

We computed the effective exchange $J$ between nearest neighbor (NN) copper
atoms as well as the parameters of a Hubbard model (that is $t$, $U$ and
$\delta \varepsilon$) as a function of the local ion distribution on the YBCO
Ba site. Following the above specifications for building the
  embedded fragment, the quantum part was reduced to a $\rm Cu_2O_9$ fragment
  surrounded by 6 $\rm Cu^{2+}$ and 10 $O^{2-}$ TIPs in the $\rm CuO_2$ plane,
  2 $\rm Cu\,I$ TIPs on top of the apical oxygens, 6 $\rm (Ca/La)^{(3-x)+}$
  averaged TIPs on the YBCO Y sites, and 6 either $\rm Ba^{2+}$ or $\rm
  La^{3+}$ on the YBCO Ba sites, as pictured in figure~\ref{fig:frag}.
  Distributing $\rm Ba^{2+}$ and $\rm La^{3+}$ ions in all possible manners in
  the 6 available positions next to the $\rm Cu_2O_9$ fragment, one finds 24
  non-equivalent configurations.  The probability of each distribution depends
  on the value of $x$.  According to the X-Ray structural data of
  reference~\cite{Struc}, each of these configurations is associated with a
  specific displacement of the apical oxygens of the $\rm Cu_2O_9$ fragment as
  well as a specific localization of the $\rm Ba^{2+}$ and $\rm La^{3+}$ ions,
  according to their nature. For each configuration, the atomic positions used
  in our calculations (both for the $\rm Cu_2O_9$ fragment and the TIPs) were 
  extracted from the experimental data of reference~\cite{Struc}.
The set of renormalized charges was built so that to reproduce the ``long''
range part of the Madelung potential of the crystal, averaged over ions
disorder. The renormalization factors from the formal charges were computed
along the method of reference~\cite{env2}. We imposed the nullity of all
multipolar moments up to order 8. It resulted a set of 90616 renormalized
charges.  The charges locations were set to the experimental crystallographic
positions given in reference~\cite{Struc}, without the local displacements
associated with a specific local ions configuration. The ions nominal charges
were taken as the averaged nominal charges on the crystallographic site~; that
is for the superconducting $\rm CuO_2$ layers and related apical oxygens~:
$\rm Cu\equiv 2+$, $\rm O\equiv 2-$, for the doping $\rm CuO_{y-6}$ layers~:
$\rm Cu \equiv 2+$, $\rm O\equiv \eta \times 2-$, where $\eta$ is the site
occupation probability in the AFM parent system, for the $\rm Ba/La$ and $\rm
Ca/La$ layers $2.125+x/2$ and $3-x$.  The Madelung potential obtained from
this set of renormalized charges exhibited an error smaller than 1~mev on any
site of the quantum part.  We also ran test calculations in order to verify
that (i) the charge modifications associated with the doping from the AFM
parent, (ii) as well as the actual ion distribution in those far away layers,
do not affect the results in any significant way (error bar 2-3~meV).

We computed the $J$, $t$, $U$ and $\delta \varepsilon$ parameters for the 24
distributions, for $x=0.1$ and $x=0.4$ and the hole doping parameter $y$
associated with the maximal critical temperature given in
reference~\cite{Struc} ---~that is $y=7.158$, $T_c=52.6\,\rm K$ for $x=0.1$
and $y=7.174$, $T_c=80.3\,\rm K$ for $x=0.4$.

\begin{figure}[h] \centerline{
\resizebox{6cm}{!}{\includegraphics{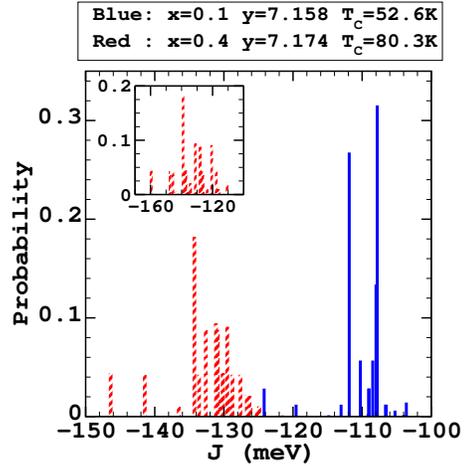}}}
\caption{(Color online) Probability of effective exchange
  values. Antiferromagnetic exchange $J$ are taken negative. The inset shows
  the probability of occurrence for effective exchange values when the atomic
  displacements (induced by the nature of the surrounding Ba site ions) of the
  apical oxygens are ignored.  Solid blue lines are associated with $\rm
  (Ca_{0.1}La_{0.9})(Ba_{1.65}La_{0.35})Cu_3O_{7.158}$ and dashed red lines
  with $\rm (Ca_{0.4}La_{0.6})(Ba_{1.35}La_{0.65})Cu_3O_{7.174}$.}
\label{fig:J}
\vspace*{-1eM}
\end{figure}
\begin{figure}[h!]\centerline{
\resizebox{!}{5.5cm}{\includegraphics{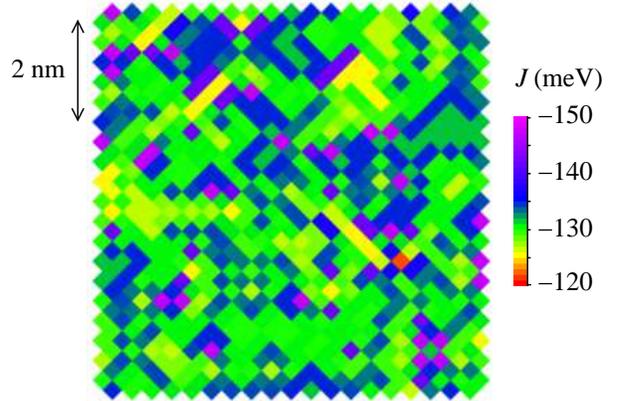}}}
\caption{(Color online) Real-space map of magnetic exchange obtained for a
  random distribution of the Ba and La ions in the YBCO Ba sites in the $\rm
  (Ca_{0.4}La_{0.6})(Ba_{1.35}La_{0.65})Cu_3O_{7.174}$ compound.}
\label{fig:Jcarte}
\end{figure}

Figure~\ref{fig:J} presents the probability of each $J$ value occurrence and
figure~\ref{fig:Jcarte} displays the resulting real-space map of magnetic
exchange obtained for a random distribution of the Ba and La ions in the YBCO
Ba sites.  One sees immediately that the $J$ values are spread over a range
wider than 20~meV for each system despite the fact the Cu--O--Cu buckling
angle does not depend on the $\rm Ba\,/\,La_{Ba}$ distribution. Indeed, the
buckling angle is only system dependant. For a given compound, the $J$
distribution is thus independent of it, while this is the contrary for the
average exchange value, $\bar J$.  One finds, for the two studied compounds, 
$\overline{J_{x=0.1}}=-110\,\rm meV$ for a buckling angle of
$\theta=167.2\ensuremath{^\circ}$ and $\overline{J_{x=0.4}}=-132\,\rm meV$ for
$\theta=170.0\ensuremath{^\circ}$. Let us note that the variation of $\bar J$
with $\theta$ is as expected, that is  the super-exchange term increases
with the overlap between the magnetic Cu $3d_{x^2-y^2}$ orbitals and the
bridging in-plane oxygen $2p$ orbitals (i.e. with $\cos^2{(\theta-\pi)}$).

Let us now analyze the possible origin of the fluctuations around these
average values. The buckling angle and the embedding can be eliminated since
they are unique for each compound. The only remaining degrees of freedom are
related with the local $\rm Ba\,/\, La_{Ba}$ distributions. One can think of
i) the modification of the apical oxygen ($\rm O_c$) localization as a
function of the local $\rm Ba\,/\, La_{Ba}$ distribution, and ii) the
electrostatic potential fluctuations induced by the different $\rm
Ba^{2+}\,/\, La_{Ba}^{3+}$ charges.  The relative influence of these two
degrees of freedom can easily be determined by computing the $J$ values for
each $\rm Ba\,/\, La_{Ba}$ distributions but with undistorted geometries (see
inset of figure~\ref{fig:J}). Doing so, one sees that i) the $J$ values still
present a large dispersion and ii) that  the dispersion range is much
larger (nearly double) than when the apical oxygen displacements are taken
into account. It results that contrarily to what is usually admitted, the
electrostatic effects play an important role in the effective exchange
values. In fact, one of the mayor role of the atomic position relaxation is to
moderate this effect. One can thus expect that the actual spatial fluctuations
of $J$ is even larger than the ones calculated in this work and displayed in
figure~\ref{fig:Jcarte}. Indeed, not only the electrostatic disorder induced
by the Ca\,/\,La distribution on the YBCO Y site has not been explicitely
taken into account in the present calculations, but in addition there are not
any atomic displacement associated with these Ca\,/\,La distributions that
could soften their electrostatic effects.

\begin{figure}[h] \centerline{
\resizebox{6cm}{!}{\includegraphics{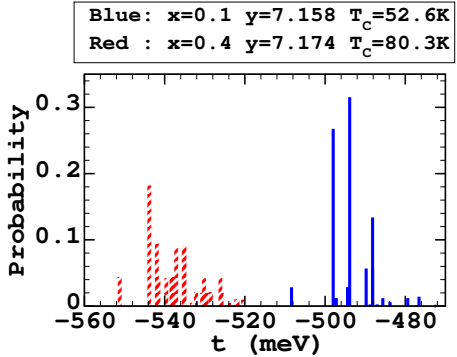}}}
\caption{(Color online) Probability of effective transfer integrals values.
  Solid blue lines are associated with $\rm
  (Ca_{0.1}La_{0.9})(Ba_{1.65}La_{0.35})Cu_3O_{7.158}$ and dashed red lines
  with $\rm (Ca_{0.4}La_{0.6})(Ba_{1.35}La_{0.65})Cu_3O_{7.174}$.}
\label{fig:t}
\end{figure}
Figure~\ref{fig:t} displays the probability of occurrence of the effective
hopping values between NN copper atoms.  The average transfer for the two
compounds $\bar t$ is found to be $\overline{t_{x=0.1}}= -493\,\rm meV$ and
$\overline{t_{x=0.4}}=-537\,\rm meV$, while the ranges of variation are
respectively $36\,\rm meV$ and $37\,\rm meV$.  Basically, the effective
transfer integrals behave in a similar way as the effective exchange, both as
a function of the buckling angle and of the local electrostatic fluctuations.
As far as the on-site repulsion $U$ is concerned, the system average values
are very similar with $\overline{U_{x=0.1}}=8.9\,\rm eV$ and
$\overline{U_{x=0.4}}=8.8\,\rm eV$, and the spatial fluctuations very small in
relative values with $(\Delta U/U)_{x=0.1}=0.07$ and $(\Delta
U/U)_{x=0.4}=0.06$.  Let us note that the $J$ fluctuations nicely agree
with their perturbative evaluations from $t$ and $U$ fluctuations, that is
$\Delta J \simeq -8 (t/U)\Delta t + 4 (t/U)^2\Delta U$.

\begin{figure}[h]\centerline{
\resizebox{6cm}{!}{\includegraphics{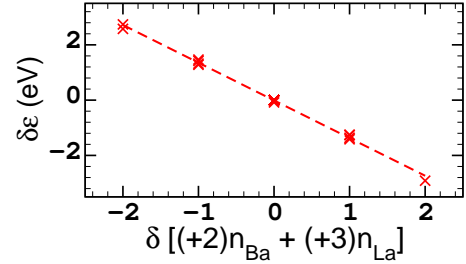}}}
\caption{Magnetic orbital energy difference between NN sites as a function of
  the charge difference associated with the coppers first neighbors YBCO
  Ba sites. Crosses: computed points, dashed line: linear fit.}
\label{fig:delta_ch}
\end{figure}
\begin{figure}[h!]\centerline{
\resizebox{8cm}{!}{\includegraphics{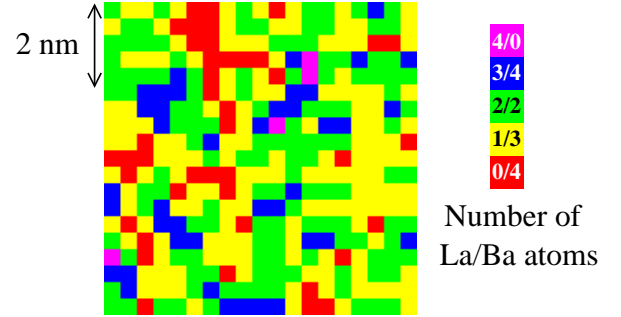}}}
\caption{(Color online) Real-space map of the number of Ba\,/\,La ions
  surrounding a given copper site (or equivalently of the magnetic orbital
  energy) for a random distribution of the Ba\,/\,La atoms in the YBCO
  Ba sites.}
\label{fig:charge}
\end{figure}
From the ab initio calculations (excitation spectrum and associated wave
functions) one can also derive the difference between nearest neighbor
magnetic orbital energies~: $\delta\varepsilon$. As for the exchange, transfer
and on-site Coulomb repulsion the effective energies of the magnetic orbitals
do strongly fluctuate in real space. In fact, the orbital energy difference
between NN sites is proportional to the difference in the number of Ba\,/\,La
ions surrounding each copper atom (see figure~\ref{fig:delta_ch}). It comes an
energy stabilization of 1.4\,eV per additional $\rm Ba \rightarrow La$
substitution. One can thus expect an associated real-space electronic density
fluctuation as a function of the local ionic configuration. Indeed, our
calculations show a correlated fluctuation in the copper atoms population of
about $0.06 \bar e$. Figure~\ref{fig:charge} pictures such an expected
fluctuation for a random distribution of the Ba\,/\,La atoms in the YBCO
Ba sites. The typical scale of the magnetic orbital energies fluctuations and
thus of the expected density of states is of the order of a few
nanometers. 

In summary our calculations showed that the extrinsic inhomogeneities
associated with material doping is responsible for both magnetic exchange and
on-site orbital energies strong spatial fluctuations. In both cases, the
fluctuations are of a typical scale of a few nanometers. 
First, the computed magnetic orbital energy fluctuations induce an associated
fluctuation of the copper density of states and positively correlate (similar
typical size and pattern) with the observed LDOS maps as picture in figure~1b
of reference~\cite{Pan01}. 
Second, the strong relation between magnetism and superconductivity in
high-$T_c$ cuprates~\cite{Tremblay08,Anderson04,Xino,Keren06} induces us to
compare our exchange fluctuations with the gap inhomogeneities observed in STS
experiments~\cite{Cren00,Pan01}. Despite the fact that it relates to different
compounds, the typical size of doping inhomogeneities is similar and
figure~\ref{fig:Jcarte} of the present work compares positively with, for
instance, figure~1 of reference \cite{Pan02}, in all aspects except for the
amplitude of the inhomogeneities. However as discussed above, the
electrostatic disorder of the YCBO Y sites was not taken into account and
should strongly increases this amplitude.
As a matter of example, we computed the $J$ modification associated with a
strongly dissymmetric distribution of the Ca\,/\,La ions on the YCBO Y sites
and found that the exchange integral is being changed from -146\,meV to
-98\,meV.

In the $\rm Bi_2Sr_2CaCu_2O_{8+\delta}$ compounds, the chemical and structural
inhomogeneities located close to the superconducting layers are, of course,
not due to the counter-ions (as in the present compounds) but rather to the
non stoichiometric oxygen atoms. Our results clearly show that this is i) the
fluctuations of the apical oxygen location and ii) the local electrostatic
environment of the superconducting copper plane that are responsible of the
exchange and orbital energy fluctuations.  The chemical origin of these
structural and electrostatic effects are system dependent, related to doping
counter-ions in our case, related to the non stoichiometric oxygen in Bi2212
compounds.
In any case our results support the conclusions
of McElroy {\it et al}~\cite{McElroy05} of an extrinsic (chemical and
structural) origin of the gap inhomogeneities and associated critical
temperature.

\acknowledgments The authors thank Daniel Maynau for providing them with
the CASDI suite of programs. These calculations where done using the
CNRS IDRIS computational facilities under project n$^\circ$1842 and
the CRIHAN computational facilities under project n$^\circ$2007013.


\end{document}